\documentclass{aa}

\usepackage[dvips]{graphicx}
\usepackage{txfonts}
\usepackage{natbib}

\begin{document}

\title{Gravity darkening in rotating stars}

\subtitle{}

\author{F. Espinosa Lara\inst{1,2}
\and M. Rieutord\inst{1,2}} 

\institute{Universit\'e de Toulouse; UPS-OMP; IRAP; Toulouse, France
\and CNRS; IRAP; 14, avenue Edouard Belin, F-31400 Toulouse, France}

\date{\today}

\abstract{Interpretation of interferometric observations of rapidly
rotating stars requires a good model of their surface effective
temperature. Until now, laws of the form $T_\mathrm{eff}\propto
g_\mathrm{eff}^\beta$ have been used, but they are only valid for
slowly rotating stars.}{We propose a simple model that can describe
the latitudinal variations in the flux of rotating stars at any rotation
rate.}{This model assumes that the energy flux is a divergence-free vector that is
antiparallel to the effective gravity.}{When mass distribution can be
described by a Roche model, the latitudinal variations in the effective
temperature only depend on a single parameter, namely the ratio of the
equatorial velocity to the Keplerian velocity. We validate this model by
comparing its predictions to those of the most realistic two-dimensional
models of rotating stars issued from the ESTER code. The agreement is
very good, as it is with the observations of two rapidly rotating stars,
$\alpha$ Aql and $\alpha$ Leo.}{We suggest that as long as a gray
atmosphere can be accepted, the inversion of data on flux distribution
coming from interferometric observations of rotating stars uses such a
model, which has just one free parameter.}


\maketitle

\section{Introduction}

The recent development of long-baseline optical/infrared interferometry
has allowed direct observation of gravity darkening at the surface of
some rapidly rotating stars. This phenomenon has been known since the work of
\citet{vonzeipel}, who noticed that the radiative flux is proportional to
the effective gravity $g_\mathrm{eff}$ in a barotropic star (i.e. when the
pressure only depends on the density). Thus, the effective temperature
$T_\mathrm{eff}$ varies with latitude, following $T_\mathrm{eff}\propto
g_\mathrm{eff}^{1/4}$. This is known as the von Zeipel law.

Barotropicity, however, is a strong hypothesis, and is actually incompatible
with a radiative zone in solid body rotation. This problem was
noticed by \citet{eddington_vonzeipel} and has led to many discussions over
subsequent decades \citep[see][for a review]{R05}. Now, the recent
interferometric observations of some nearby rapidly rotating stars by
\citet{McAlister2005}, \citet{Aufdenberg2006}, \citet{vanbelle2006}, and
\citet{Zhao2009} have shown that gravity darkening is not well represented by
von Zeipel's law, which seems to overestimate the temperature  difference
between pole and equator. This conclusion is usually expressed using a
power law, $T_\mathrm{eff}\propto g_\mathrm{eff}^\beta$, with an exponent
less than $1/4$. This is traditionally explained by the existence of
a thin convective layer at the surface of the star, in reference to
the work of \citet{lucy67}, who found $\beta\sim0.08$ for stars with a
convective envelope.

Actually, the weaker variation in the flux with latitude, compared
to von Zeipel's law has also been noticed by \citet{Lovekin2006},
\citet{Espinosa&Rieutord2007}, and \citet{Espinosa2010} in two-dimensional
numerical models of rotating stars. This prompted us to reexamine this
question using these new models.  It leads us to a new approach that
avoids the strong assumption of von Zeipel and that is presumably able
to better represent the gravity darkening of fast rotating
stars. This paper aims at presenting this new model.

In Sect. \ref{sec_model} we explain the derivation of the new model
of gravity darkening and the associated assumptions. In the next
section, results are compared to the values of two-dimensional models
and to observations. Conclusions follow.

\section{The model}
\label{sec_model}

Before presenting our model, we briefly recall the origin of
the von Zeipel law and Lucy's exponent. In a radiative region the flux
is essentially carried by the diffusion of photons and thus correctly
described by Fourier's law,
\begin{equation}
\vec F = -\chi_r\nabla T,
\end{equation}
where $\chi_r$ is the radiative heat conductivity. If the star is
assumed to be barotropic, density and temperature (and thus $\chi_r$) are
only functions of the total potential $\Phi$ (gravitational plus centrifugal). Hence,
\begin{equation} 
\vec F = -\chi_r(\Phi)\nabla T(\Phi) = -\chi_r(\Phi)T'(\Phi)\nabla\Phi= \chi_r(\Phi)T'(\Phi)\vec g_\mathrm{eff}.
\end{equation}

If we define the surface of the star as a place of constant given optical
depth, we see that this model implies latitudinal variations in the flux
that strictly follow those of the effective gravity. Using the effective
temperature, we recover $T_\mathrm{eff}\propto g_\mathrm{eff}^{1/4}$.

Lucy's approach is based on a first-order development of the dependence between $T_\mathrm{eff}$ and
$g_\mathrm{eff}$ in a convective envelope. Since the convective flux
is almost orthogonal to isentropic surfaces, $\log
T_\mathrm{eff}$ and $\log g_\mathrm{eff}$ are related on such a surface. Deep enough
where the entropy is almost the same everywhere, $s(\log T_\mathrm{eff},
\log g_\mathrm{eff})=s_0$, where $s_0$ is the specific entropy in the
bulk of  the convection zone.  Assuming that $T_\mathrm{eff}\propto
g_\mathrm{eff}^\beta$ at first order, one finds that

\begin{equation} 
\beta = -\frac{\partial \ln s/\partial \ln g_\mathrm{eff}}{\partial\ln s/\partial\ln T_\mathrm{eff}}
\end{equation}
at zero rotation. \citet{lucy67} computed this exponent for various
stellar models and found that is was weakly dependent on the mass,
radius, luminosity, or chemical abundance of the model. The mean value
$\beta\sim 0.08$ was thus proposed.

From the foregoing recap, we should remember that both approaches are
based on assumptions that require a small deviation from sphericity,
and thus they demand slow rotation.

\subsection{Hypothesis for a new model}

It is quite clear from observations that slow rotation is not
appropriate when dealing with observed gravity darkenings. We recall
that Altair ($\alpha$ Aql), a favorite target of interferometry
\citep{DKJVONA05,petersonetal06a,Monnier2007} is thought to be rotating close to
90\% of the breakup angular velocity. Thus any modeling of its gravity
darkening should be able to deal with rapid rotation.

\begin{figure}
\resizebox{\hsize}{!}{\includegraphics{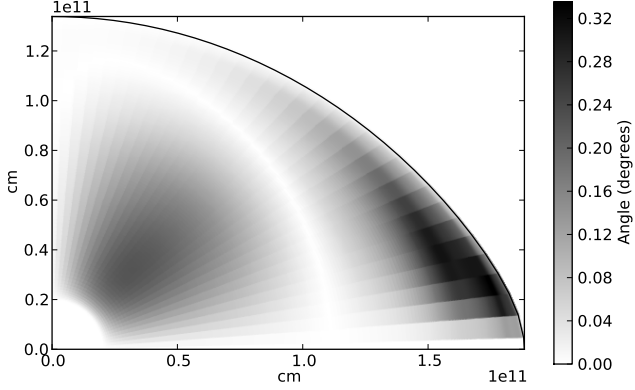}}
\caption{Meridional cut representing the angle (in degrees) formed by
the pressure gradient and the energy flux in a 3 M$_\odot$ star rotating
at 90\% of its Keplerian velocity at the equator calculated using the
ESTER code.} \label{fig_angle}
\end{figure}

We propose to assume that the flux in the envelope of a rotating star
is very close to

\begin{equation}
\label{hyp1}
\vec F = -f(r,\theta)\vec g_\mathrm{eff}.
\end{equation}
Here, $f$ is some function of the position to be determined where $r$, $\theta$,
and $\varphi$ are the spherical coordinates.
In doing so, we hypothesize that the energy flux vector is almost
antiparallel to the effective gravity. This is justified in a convective
envelope where heat transport is essentially in the vertical direction,
following buoyancy. In radiative regions, the flux is antiparallel to
the temperature gradient, which is slightly off the potential gradient
because of baroclinicity. Fortunately, the angle between the two vectors
remains very small, even in the case of extreme rotation. As shown by
Fig. \ref{fig_angle}, a two-dimensional stellar model with rotation
at 90\% of the break-up velocity gives an angle between $\nabla T$ and
$\vec g_\mathrm{eff}$ of less than half a degree (i.e. less than 10$^{-2}$
rd).
Comparing our model to that of von Zeipel, we see that it does not imply
that latitudinal variations of the flux are the same as those of the
effective gravity or, equivalently, it allows the ratio $F/g_\mathrm{eff}$
to vary with latitude.

In the envelope of a star, where no heat is generated, we have
\begin{equation}
\nabla\cdot\vec F = 0
\end{equation}
or
\begin{equation} 
\label{diff_eq}
\vec g_\mathrm{eff}\cdot\nabla f + f\nabla\cdot\vec g_\mathrm{eff} = 0.
\end{equation}
This equation is completed by an appropriate boundary condition (see
below). Presently, we need to note that the solution only depends on the
effective potential, that is, on the mass distribution. Such a  quantity
is generally not known, but we may observe that rapidly rotating stars
are usually massive or intermediately massive, hence rather centrally
condensed. We therefore chose the Roche model to represent the mass
distribution. Its simplicity is also attractive. We thus set
\begin{equation}
\label{geff_roche}
\vec g_\mathrm{eff}=\left(-\frac{GM}{r^2}+\Omega^2r\sin^2\theta\right)\vec e_r
+\Omega^2r\sin\theta\cos\theta\vec e_\theta
\end{equation}
where $\vec e_r$ and $\vec e_\theta$ are radial unit vectors associated with
the spherical coordinates.

According to this model
\begin{equation}
\lim_{r\rightarrow 0} f(r,\theta) =\frac{L}{4\pi GM}=\eta
\end{equation}
where $L$ is the luminosity of the star, $M$ its mass, and $\eta$ is
actually the constant that scales the function $f$. We therefore write

\begin{equation}
f(r,\theta) = \eta F_\omega(\tilde r,\theta) 
\end{equation}
where
\begin{equation} 
\label{omega}
\omega = \Omega\sqrt{\frac{R_e^3}{GM}}=\frac{\Omega}{\Omega_k}
\end{equation}
is the nondimensional measure of the rotation rate where $\Omega_k$ is
the Keplerian angular velocity at the equator, and $\tilde r=r/R_e$ the
dimensionless radial coordinate scaled by the equatorial radius $R_e$.
We see that in this model the latitude dependence of the flux is
controlled by a single parameter $\omega$. It is therefore as easy to
use as the previous models of von Zeipel or Lucy, but it is likely to
be much more robust at high rotation.

\subsection{Solution of the new model}

For the Roche model $\nabla\cdot\vec g_\mathrm{eff}=2\Omega^2$, so the
dimensionless form of Eq. (\ref{diff_eq}) can be written as

\begin{equation}
\label{diff_eq_dless}
\left(\frac{1}{\omega^2\tilde r^2}-\tilde r\sin^2\theta\right)
\frac{\partial F_\omega}{\partial \tilde r}
-\sin\theta\cos\theta\frac{\partial F_\omega}{\partial \theta}
=2F_\omega
\end{equation}
with the boundary condition
\begin{equation}
\label{bcond_dless}
F_\omega(\tilde r=0,\theta)=1.
\end{equation}
This linear, first-order, and homogeneous partial differential equation can be solved 
by using a change of variable based on the method
of characteristics. Along a characteristic curve
of (\ref{diff_eq_dless}) we have

\begin{equation}
\label{characteristic}
\frac{\mathrm{d}\tilde r}{\frac{1}{\omega^2\tilde r^2}-\tilde r\sin^2\theta}=
\frac{\mathrm{d}\theta}{-\sin\theta\cos\theta}=
\frac{\mathrm{d}F_\omega}{2F_\omega}.
\end{equation}
We can multiply the first and second terms by an arbitrary function $G(\tilde r,\theta)$
\begin{equation}
G(\tilde r,\theta)\frac{\mathrm{d}\tilde r}{\frac{1}{\omega^2\tilde r^2}
-\tilde r\sin^2\theta}=
G(\tilde r,\theta)\frac{\mathrm{d}\theta}{-\sin\theta\cos\theta},
\end{equation}
which leads to
\begin{equation}
\label{dtau}
G(\tilde r,\theta)\left[\sin\theta\cos\theta\mathrm{d}\tilde r+\left(\frac{1}{\omega^2\tilde r^2}
-\tilde r\sin^2\theta\right)\mathrm{d}\theta\right]=0=\mathrm{d}\tau,
\end{equation}
where we have defined a new variable $\tau$ that will be constant along a characteristic curve.

Without loss of generality we choose\footnote{The 
function $G(\tilde r,\theta)$ is in fact an
integrating factor that should be chosen so that Eq. (\ref{dtau})
is an exact differential, namely that $\frac{\partial}{\partial\tilde
r}\left(\frac{\partial\tau}{\partial\theta}\right)=
\frac{\partial}{\partial\theta}\left(\frac{\partial\tau}{\partial\tilde
r}\right)$; then a solution $\tau(\tilde r,\theta)$ exists. This form
is not unique since any function of $\tau$ will also be a solution.}

\begin{equation}
G(\tilde r,\theta)=\frac{\omega^2\tilde r^2\cos^2\theta}{\sin\theta}.
\end{equation}
This leads, via the chain rule
$\mathrm{d}\tau=\frac{\partial\tau}{\partial \tilde r}\mathrm{d}\tilde r
+\frac{\partial\tau}{\partial \theta}\mathrm{d}\theta$,
to a system of differential equations to calculate $\tau$
\begin{equation}
\left\{
\begin{array}{l}
\displaystyle \frac{\partial\tau}{\partial\tilde r}=\omega^2\tilde r^2\cos^3\theta\\
\displaystyle \frac{\partial\tau}{\partial\theta}=
\frac{\cos^2\theta}{\sin\theta}-\omega^2\tilde r^3\sin\theta\cos^2\theta.
\end{array}
\right. 
\label{taueq}
\end{equation}
From the first equation we obtain
\begin{equation}
\label{tau0}
\tau=\frac{1}{3}\omega^2\tilde r^3\cos^3\theta+h(\theta)
\end{equation}
with $h(\theta)$ an unknown function. Using (\ref{tau0}) in the
second equation of (\ref{taueq}), we find

\begin{equation}
\label{dhdt}
\frac{\mathrm{d}h}{\mathrm{d}\theta}=\frac{\cos^2\theta}{\sin\theta},
\end{equation}
which can be integrated, yielding
\begin{equation}
h(\theta)=\cos\theta+\ln\tan\frac{\theta}{2}
\end{equation}
so that (\ref{tau0}) becomes
\begin{equation}
\tau=\frac{1}{3}\omega^2\tilde r^3\cos^3\theta+\cos\theta+\ln\tan\frac{\theta}{2}.
\end{equation}
In fact, the curves $\tau=\mathrm{const.}$ are the streamlines of
the flux $\vec F$, hence also those of the effective gravity $\vec
g_\mathrm{eff}$ according to Eq. (\ref{hyp1}).

After making the change of variable $(\tilde r,\theta)\rightarrow(\tau,\theta)$,
the original differential equation (\ref{diff_eq_dless}) simplifies to
(also from Eq. (\ref{characteristic}))

\begin{equation}
-\sin\theta\cos\theta\left(\frac{\partial F_\omega}{\partial \theta}\right)_\tau
=2F_\omega,
\end{equation}
which can be solved to get the general solution
\begin{equation}
F_\omega=\frac{\psi(\tau)}{\tan^2\theta}.
\label{Fom}
\end{equation}

The function $\psi(\tau)$ can be calculated using the boundary condition
(\ref{bcond_dless}). We define a new variable
$\vartheta(\tilde r,\theta)$ such that

\begin{equation}
\label{theta0}
\cos\vartheta+\ln\tan\frac{\vartheta}{2}=
\frac{1}{3}\omega^2\tilde
r^3\cos^3\theta+\cos\theta+\ln\tan\frac{\theta}{2}=\tau,
\end{equation}
so that $\vartheta\equiv\vartheta(\tau)$. For a given value of $\tau$,
$\vartheta$ is the corresponding value of $\theta$ when $\tilde r\rightarrow0$.
From Eq. (\ref{Fom}) we easily find that

\begin{equation}
\psi(\tau)=\tan^2\vartheta
\end{equation}
and
\begin{equation}
F_\omega=\frac{\tan^2\vartheta}{\tan^2\theta}.
\end{equation}
This expression can be used to calculate the energy flux everywhere in
the star. However, we may note singularities at $\theta=0$ and
$\theta=\pi/2$, i.e. at the pole and the equator of the star. Actually,
these points are not real singularities since the direct resolution of
Eq. (\ref{diff_eq_dless}) at these points gives

\begin{equation}
\label{F_pole}
F_\omega(\theta=0)=\mathrm{e}^{\frac{2}{3}\omega^2\tilde r^3}
\end{equation}
and
\begin{equation}
\label{F_eq}
F_\omega(\theta=\pi/2)=\left(1-\omega^2\tilde r^3\right)^{-2/3}
\end{equation}
using the boundary condition (\ref{bcond_dless}).

\subsection{The effective temperature profile}\label{sec_prof}

To obtain the latitudinal profile of $T_\mathrm{eff}$ at the surface
of the star, one must proceed as follows.
Once $\omega$ is fixed, the shape of the surface
$\tilde r(\theta)$ is given by the Roche potential
\begin{equation}
\frac{GM}{r}+\frac{1}{2}\Omega^2r^2\sin^2\theta=\mathrm{const.}
\end{equation}
Using Eq. (\ref{omega}), it follows that $\tilde r$ is the solution of

\begin{equation}
\frac{1}{\omega^2\tilde r} + \frac{1}{2}\tilde r^2\sin^2\theta =
\frac{1}{\omega^2}+ \frac{1}{2}
\end{equation}
for a given $\theta$.
Thus the value of $\vartheta$ at each point on the surface ($\tilde
r,\theta$) can be calculated using (\ref{theta0}). This equation
is efficiently solved with an iterative method like the Newton one. 
Then we have

\begin{eqnarray}
T_\mathrm{eff}&=&\left(\frac{F}{\sigma}\right)^{1/4}
=\left(\frac{L}{4\pi\sigma GM}\right)^{1/4}\!\!\!\!\!\sqrt{\frac{\tan\vartheta}{\tan\theta}}
g_\mathrm{eff}^{1/4} \nonumber\\
&=&
\left(\frac{L}{4\pi\sigma R_e^2}\right)^{1/4}\!\! \left( \frac{1}{\tilde r^4} +
\omega^4\tilde r^2\sin^2\theta-\frac{2\omega^2\sin^2\theta}{\tilde
r}\right)^{1/8}\!\!\!\!\!\!\sqrt{\frac{\tan\vartheta}{\tan\theta}},
\label{big}
\end{eqnarray}
which shows that the temperature profile depends on a single parameter
$\omega$. As a consequence the ratio of equatorial to polar effective temperature
is, when using (\ref{F_pole}) and (\ref{F_eq}), given by
\begin{equation}
\frac{T_\mathrm{eff}^e}{T_\mathrm{eff}^p}=
\sqrt{\frac{2}{2+\omega^2}}(1-\omega^2)^{1/12}
\exp\left(-\frac{4}{3}\frac{\omega^2}{(2+\omega^2)^3}\right).
\end{equation}

Finally we note that von Zeipel's law is recovered at slow rotation,
since in this case $\vartheta\simeq\theta$ and Eq. (\ref{big}) simplifies into 

\begin{equation}
 T_\mathrm{eff}=
\left(\frac{L}{4\pi\sigma GM}\right)^{1/4}
g_\mathrm{eff}^{1/4}.
\end{equation}

\begin{figure*}
\centering
\begin{tabular}{ccc}
\includegraphics[width=5.6cm]{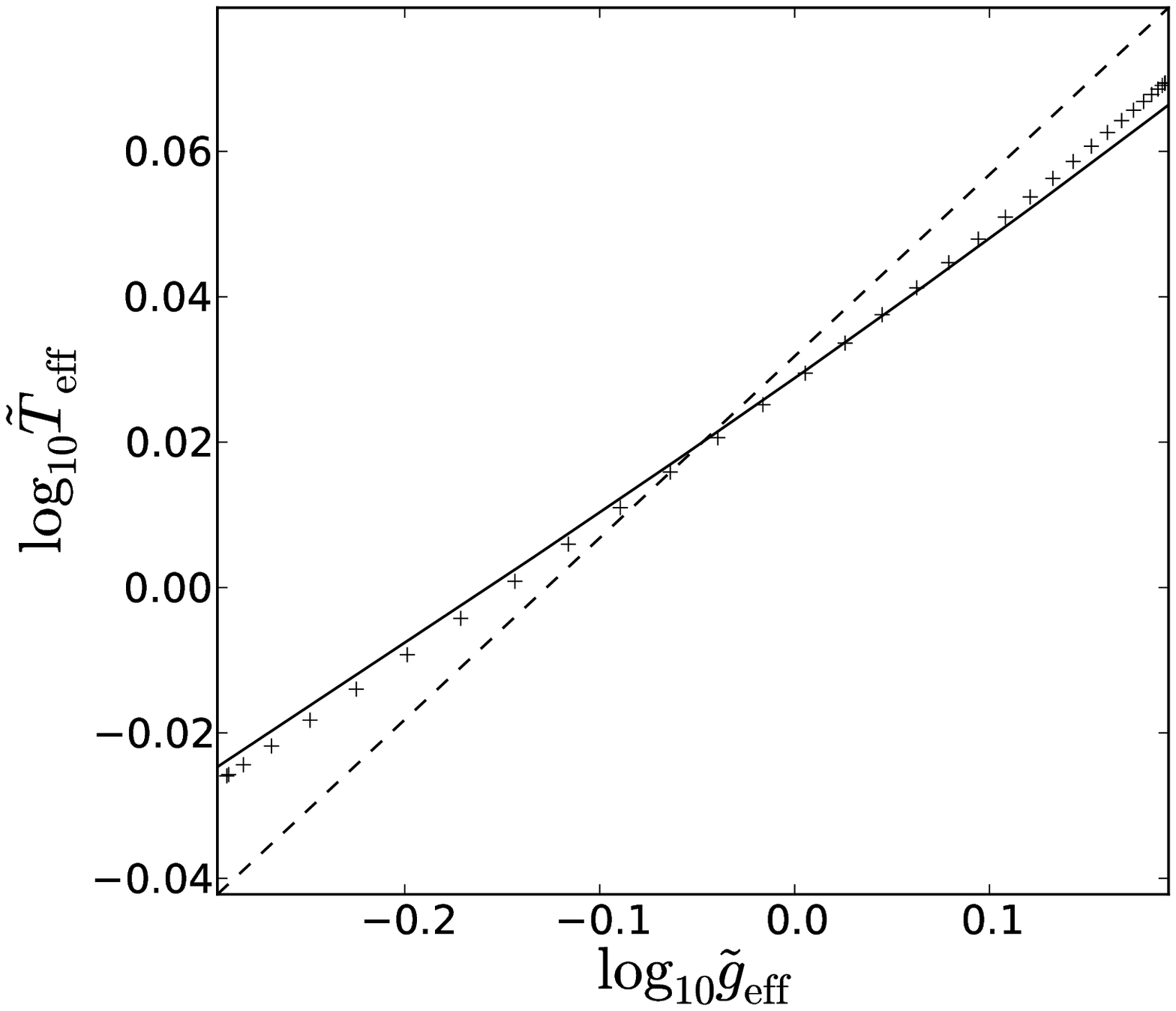}&
\includegraphics[width=5.6cm]{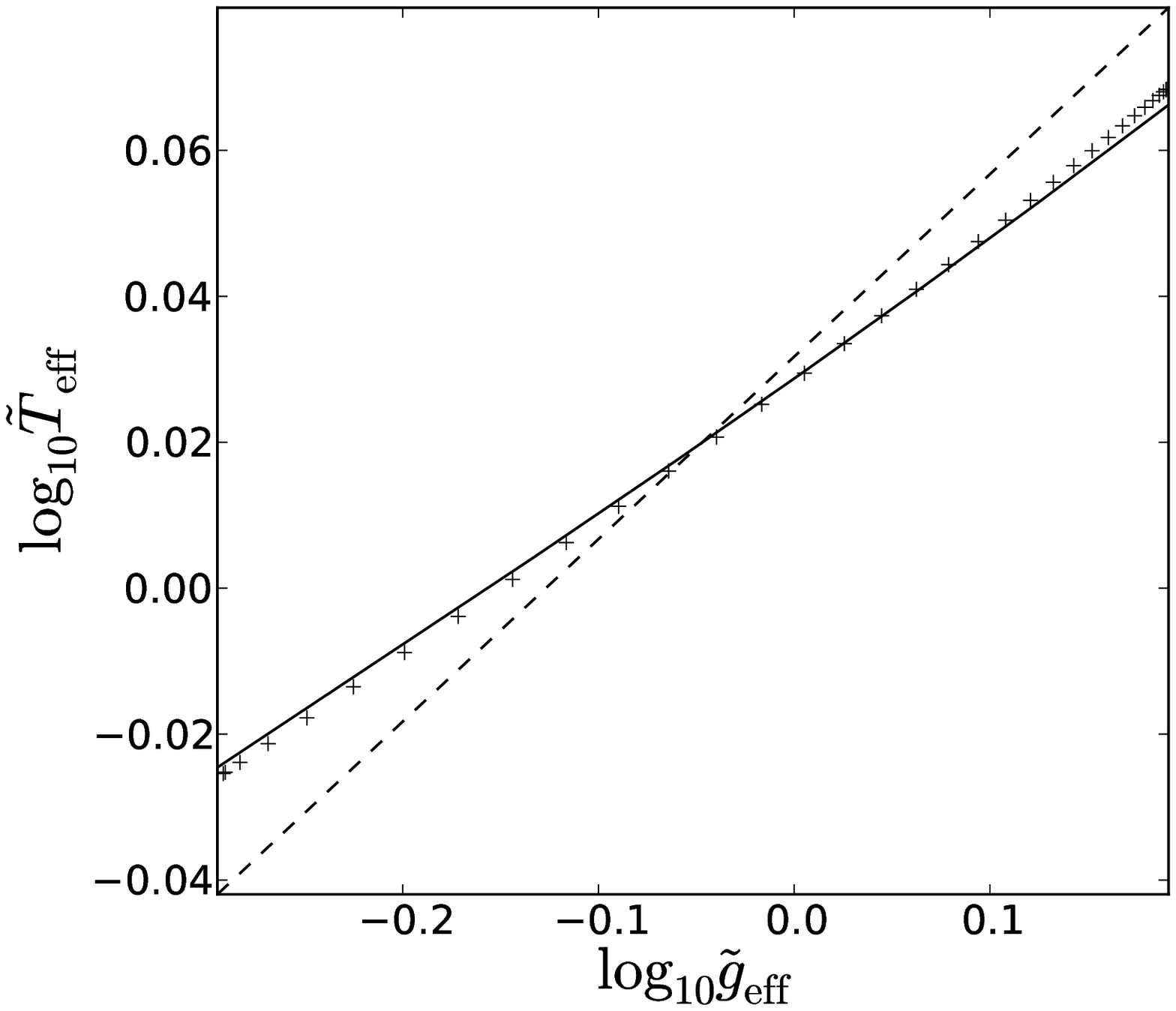}&
\includegraphics[width=5.6cm]{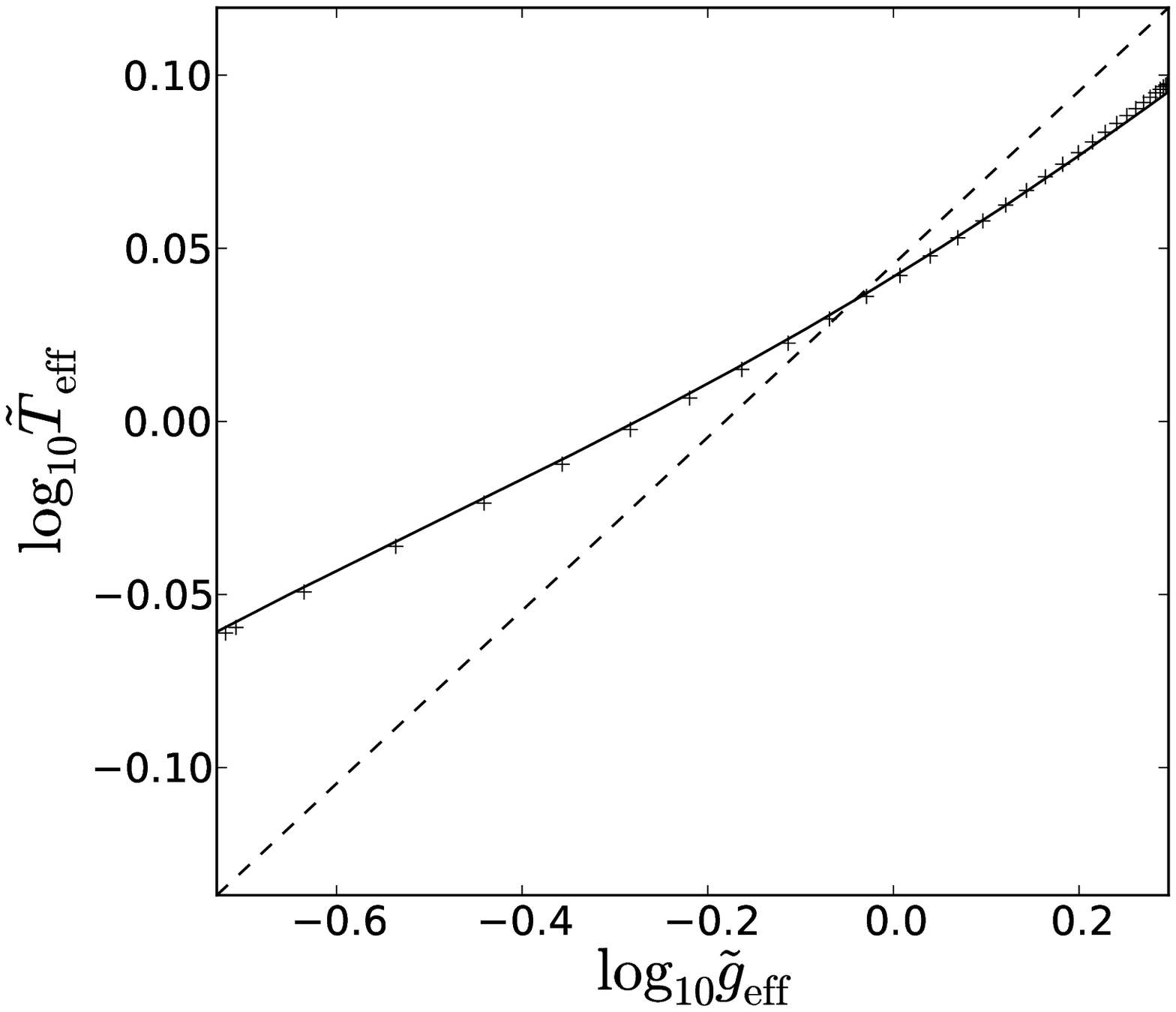}\\
\includegraphics[width=5.6cm]{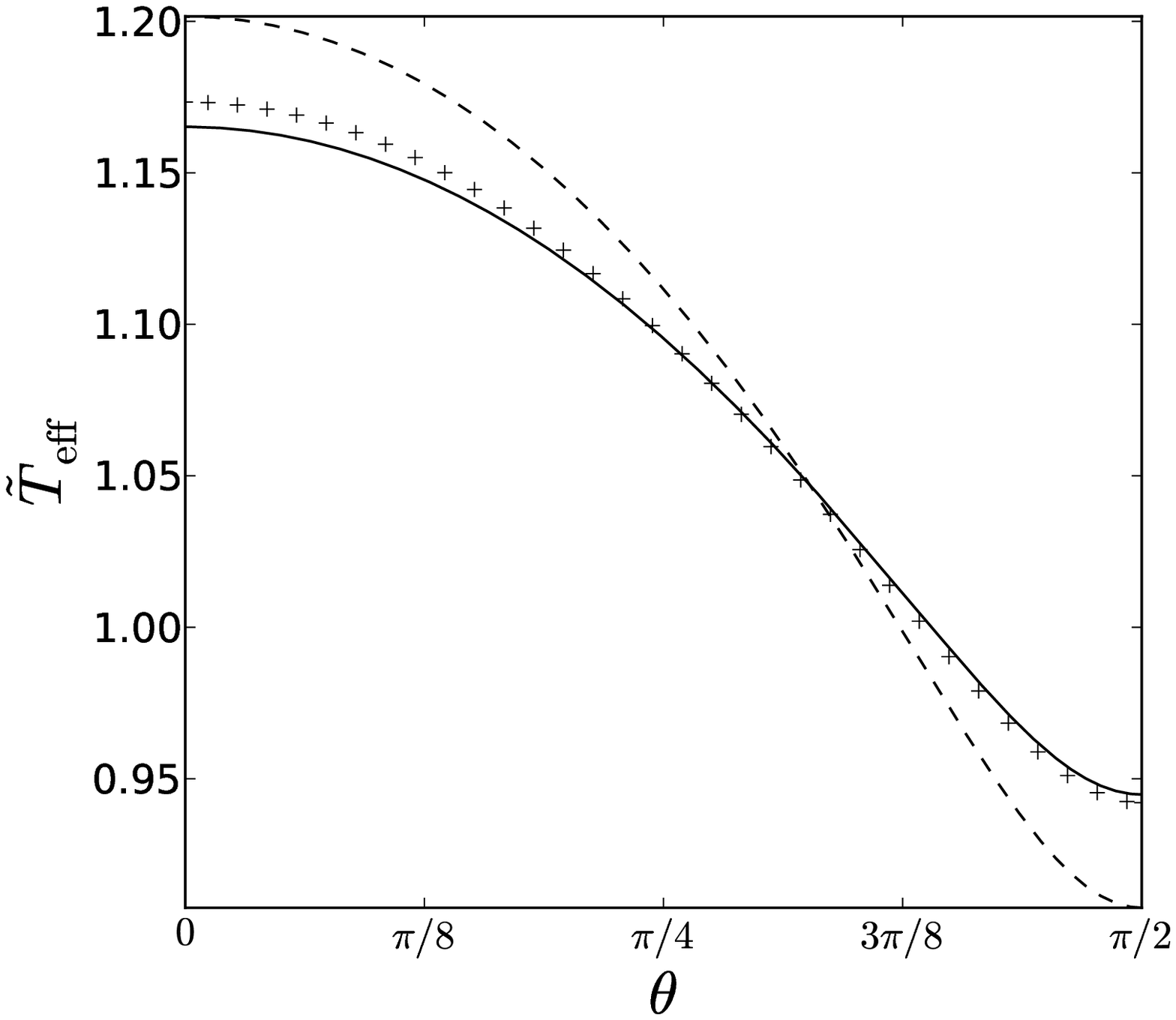}&
\includegraphics[width=5.6cm]{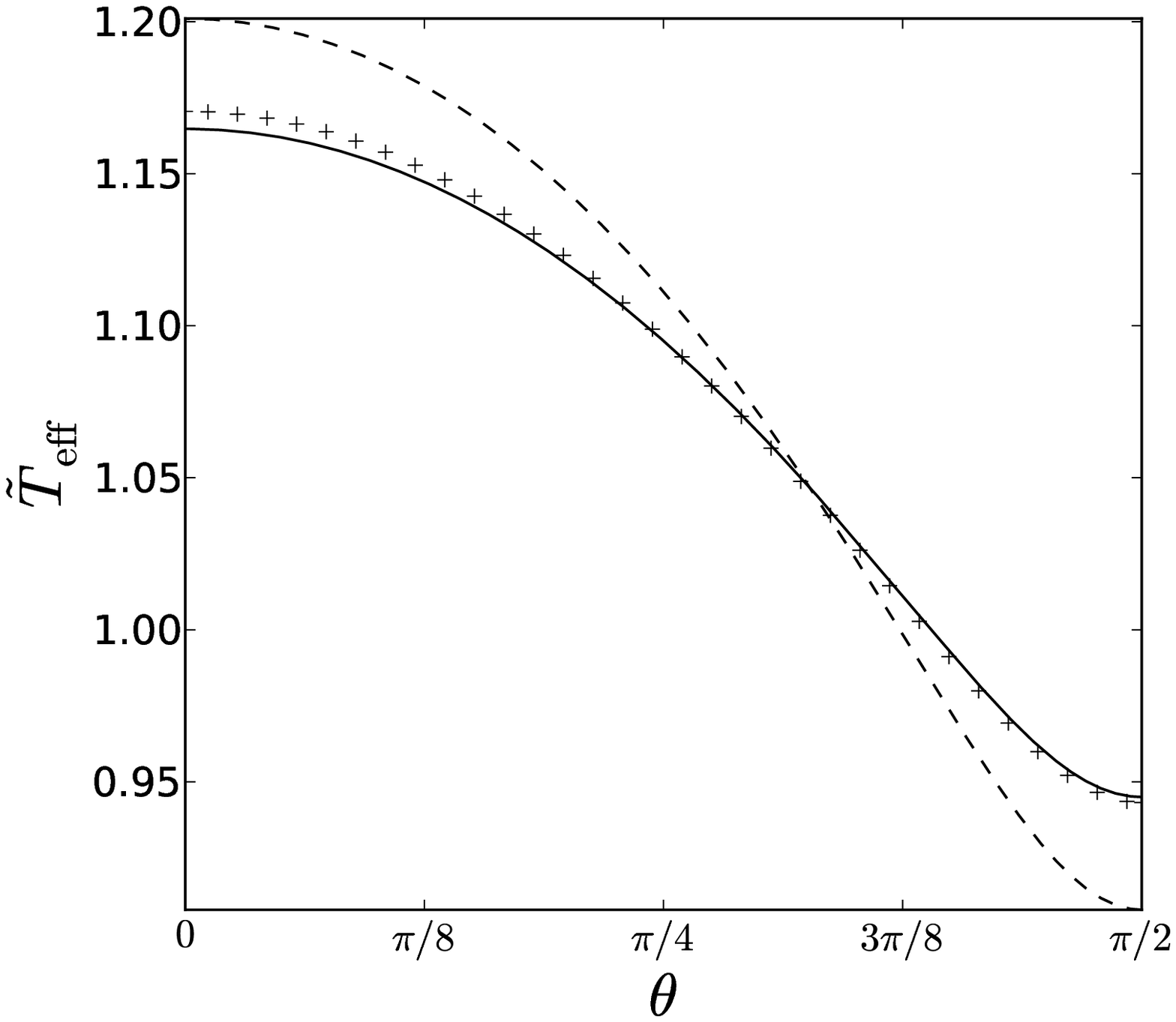}&
\includegraphics[width=5.6cm]{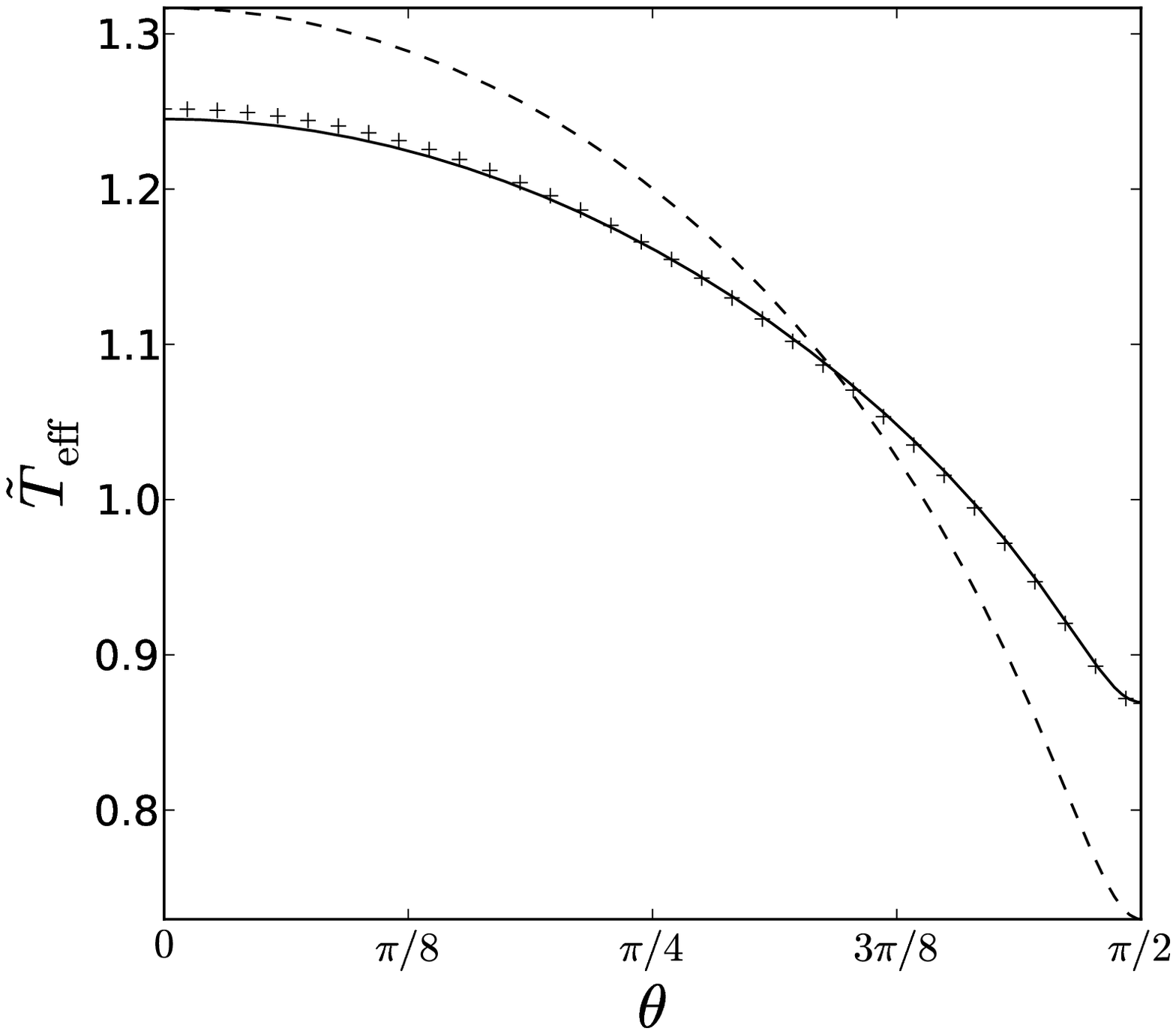}
\end{tabular}
\caption{Comparison between the gravity darkening profile calculated
using the model described in Sect. \ref{sec_prof} (solid line), using
von Zeipel's law (dashed line), and the models calculated using
the ESTER code (crosses). \emph{Left}: Star with  $M=3 M_\mathrm{\sun}$
and homogeneous composition rotating at 70\% of the Keplerian velocity
at the equator ($\Omega_k$). \emph{Center}: $M=3M_\mathrm{\sun}$,
$\Omega=0.7\Omega_k$ and with hydrogen abundance in the core $X_c=0.5X$,
with $X$ the abundance in the envelope. \emph{Right}: Same as center
but with $\Omega=0.9\Omega_k$. In the bottom row $\theta$ is the
colatitude.}
\label{fig0}
\end{figure*}

\section{Results and comparison with ESTER models}
\label{sec_results}

To validate the foregoing model of gravity darkening, we compare
its results to those of fully two-dimensional models of rotating stars
produced by the ESTER\footnote{Evolution STEllaire en Rotation} code
\citep[e.g.][]{Espinosa&Rieutord2007,Rieutord&Espinosa2009,Espinosa2010}.
These models utilize the OPAL opacities and equation of state
\citep{opal_opac,opal_eos}.  They include a convective core and a
fully radiative envelope. They are presently calculated in the limit of
vanishing viscosity, which requires a prescription for the differential
rotation. Here, we have chosen the surface rotation to be solid. The
interior rotation is then self-consistently derived.

We have computed two sets of models. The first set has homogeneous
composition with masses between 2.5~M$_\odot$ and 4~M$_\odot$ and the full
range of rotation velocities, from zero up to the breakup velocity.
The second set covers the same range of masses and velocities, but we have
set the hydrogen abundance in the core to be 50\% of the abundance in the
envelope. This results in a smaller core compared to the size of the star.

\begin{figure}
\resizebox{\hsize}{!}{\includegraphics{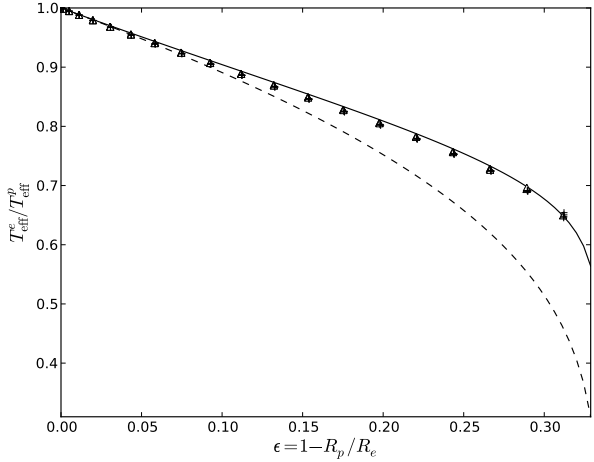}}
\caption{Ratio between equatorial and polar effective temperature as
a function of flattening $\epsilon=1-R_p/R_e$. \emph{Solid line}:
The analytical model presented in this paper.  \emph{Dashed line}: Von Zeipel's
law. \emph{Crosses}: ESTER models of set 1 ($X_c=X$).  \emph{Triangles}:
ESTER models of set 2 ($X_c=0.5X$).  }
\label{fig1}
\end{figure}

Figure \ref{fig0} shows the comparison between our model of gravity
darkening and the stellar models calculated using the ESTER code. It
represents the profile of effective temperature as a function of surface
effective gravity (top) and colatitude (bottom). The normalized values are defined
as $\tilde T_\mathrm{eff}=T_\mathrm{eff}\left(\frac{L}{4\pi\sigma
R_e^2}\right)^{-1/4}$ and $\tilde
g_\mathrm{eff}=g_\mathrm{eff}\left(\frac{GM}{R_e^2}\right)^{-1}$. For
comparison, we have also plotted the profile predicted by von Zeipel's
law. The first two columns on the left show the results for a star rotating at
70\% of the Keplerian velocity at the equator\footnote{Although
Keplerian velocity $\Omega_k=\sqrt{g_e/R_e}$ can be identified as
the critical velocity, we use the former term to avoid confusion
with other definitions of critical velocity seen in most works.
A common definition is based on a series of models (typically Roche
models) of increasing rotation rate, in which the critical velocity
$\Omega_c$ is defined as the angular velocity of the model rotating
at the break-up limit. For Roche models, $\Omega_c=\sqrt{8GM/27R_p^3}$.
These definitions are not equivalent, a star with $\Omega=0.7\Omega_k$
will have $\Omega\simeq0.93\Omega_c$.} $\Omega_k$, while the third
one corresponds to $\Omega=0.9\Omega_k$. The difference between
the first and the second columns is the hydrogen abundance in the core.
As we observe, von Zeipel's law systematically overestimates the
ratio between polar and equatorial effective temperature, while there
is good agreement between the predictions of $F_{\omega}$ and the
ESTER models. The same behavior can be seen in Fig. \ref{fig1}, which
shows the ratio between equatorial and polar effective temperature as
a function of the flattening of the star $\epsilon=1-R_p/R_e$, even at
high rotation rates.

The good agreement between the profile of the effective temperature
predicted by the new model and the output of the ESTER code is
general. However, there are some slight deviations, which are more important
for the models with homogeneous composition. We think that this may
be a consequence of the differential rotation of ESTER models, which
contain a convective core rotating faster than the envelope. This
leads to a different shape of equipotential surfaces (or isobars)
compared to the rigidly rotating Roche model that is used in the simple
model. This difference is less important for ``evolved" models, because they
have smaller cores.

\begin{figure}
\resizebox{\hsize}{!}{\includegraphics{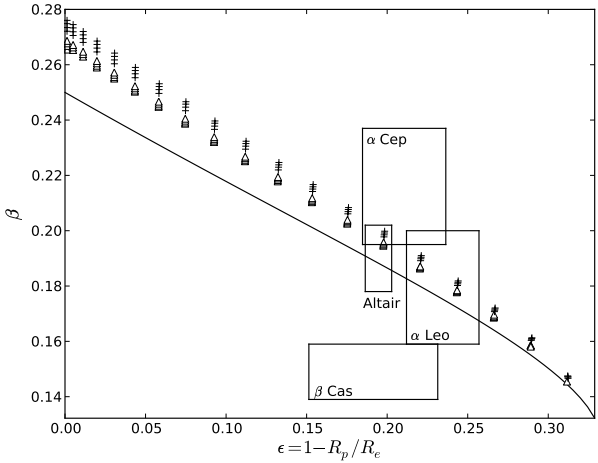}}
\caption{Gravity darkening coefficient $\beta$ as a function of
flattening. The values were calculated using a least-square fit
to the profile of effective temperature.  \emph{Solid line}: Model
presented in this paper.  \emph{Crosses}: ESTER models of set 1 ($X_c=X$).
\emph{Triangles}: ESTER models of set 2 ($X_c=0.5X$).  The corresponding
value of $\beta$ for von Zeipel's law is 0.25.  The boxes represent
the values of $\beta$, with their corresponding errors, obtained by
interferometry for several rotating stars: Altair \citep{Monnier2007},
$\alpha$ Cephei \citep{Zhao2009}, $\beta$ Cassiopeiae, and $\alpha$ Leonis
\citep{Che2011}.}
\label{fig2}
\end{figure}

The gravity darkening exponent $\beta$, derived from a generalization of
von Zeipel's law $T_\mathrm{eff}=g_\mathrm{eff}^\beta$, is commonly used
in observational work to measure the strength of gravity darkening.
In Fig. \ref{fig0} we can observe that the relation between effective
temperature and gravity is not exactly a power law. Using such a law
is therefore not quite appropriate, but since $\beta$-exponents have been
derived from observations, we feel it is worth deriving a best-fit exponent
associated with ESTER models. These are shown in Fig.~\ref{fig2}. This
plot clearly shows that the value $\beta=0.25$ given by von Zeipel's
law is appropriate only {\em in the limit of slow rotation} as expected.
In this figure we also plotted measurements of the gravity darkening
exponent $\beta$ derived by interferometry for several rapidly rotating
stars, Altair \citep{Monnier2007}, $\alpha$ Cephei \citep{Zhao2009},
$\beta$ Cassiopeiae, and $\alpha$ Leonis \citep{Che2011}.
We can see that there is good (Altair, $\alpha$ Leo) or fair ($\alpha$
Cep) agreement between the observed values and the model. The discrepancy
for $\beta$ Cas may come from its small inclination angle ($\sim20$
deg), so we see the star near pole-on, which makes determining
of its flattening more difficult and makes the results depend more on
the model used for gravity darkening and limb darkening.

\section{Conclusions}
\label{sec_conclusions}

Observing that the energy radiated by a star is produced almost entirely
in the stellar core and that most stars may be considered close to a
steady state, we have noted that the energy flux is essentially a
divergence-free vector field. We also observed from full two-dimensional
models of rotating stars that the direction of this vector is always
very close to that of the effective gravity, so it only depends on the mass
distribution. Following this picture, the physical conditions in the
outer layers can only affect the flux radiated outside
the star very slightly as long as a gray atmosphere can be assumed. Thus, we proposed
that the energy flux in the envelope of a rotating star be approximated
by $\vec F = -\frac{L}{4\pi GM}F_\omega(r,\theta)\vec g_{\rm eff}$.

We have shown how the non-dimensional function $F_\omega(r,\theta)$ can
be evaluated by assuming that mass distribution is represented by the Roche
model. We then demonstrated that the latitudinal variation of the
effective temperature only depends on a single parameter $\omega=
\sqrt{\Omega^2R_e^3/GM}$.
Such a model is very appropriate to interpreting the
interferometric observations of rotating stars since, unlike von
Zeipel's or Lucy's laws, it is valid for high rotation rates (up to breakup)
and depends only on a single parameter (the $\beta$-exponent is removed).
Adjustment of the observed surface flux would thus only require
variations in $\omega$ and $i$, the inclination of the rotation axis on
the line of sight.

As previously mentioned, this model fits the fully 2D models well using
a gray atmosphere with a rigidly rotating surface (but with interior
differential rotation). This dynamical feature of the models may not be
very realistic, so we tested a surface rotation
$\Omega(\theta)=\Omega_{\rm eq}(1-0.1\cos^2\theta)$ inspired by
observations \citep{CC07}. The difference is hardly perceptible,
therefore the proposed model of gravity darkening looks quite robust.
Future improvement of two-dimensional models will of
course be used to confirm this robustness.

\begin{acknowledgements}
The authors acknowledge the support of the French Agence Nationale de
la Recherche (ANR), under grant ESTER (ANR-09-BLAN-0140).  This work
was also supported by the Centre National de la Recherche Scientifique
(C.N.R.S., UMR 5572), through the Programme National de Physique Stellaire
(P.N.P.S.). The numerical calculations were carried out on the
CalMip machine of the `Centre Interuniversitaire de Calcul de Toulouse'
(CICT), which is gratefully acknowledged.
\end{acknowledgements}

\bibliographystyle{aa} 
\bibliography{biblio} 

\end{document}